\documentclass[pdflatex,sn-mathphys-num]{sn-jnl}% Math and Physical Sciences Numbered Reference Style
\usepackage{graphicx}%
\usepackage{multirow}%
\usepackage{amsmath,amssymb,amsfonts}%
\usepackage{amsthm}%
\usepackage{mathrsfs}%
\usepackage[title]{appendix}%
\usepackage{xcolor}%
\usepackage{textcomp}%
\usepackage{manyfoot}%
\usepackage{booktabs}%
\usepackage{algorithm}%
\usepackage{algorithmicx}%
\usepackage{algpseudocode}%
\usepackage{listings}%
\usepackage{subfigure}
\usepackage{makecell}
\usepackage{float}
\theoremstyle{thmstyleone}%
\theoremstyle{thmstyletwo}%

\theoremstyle{thmstylethree}%

\begin{document}

\title[Nuclear Structure and Shape Evolution of Nd Isotopes]{Nuclear Structure and Shape Evolution of Nd Isotopes}


\author*[1]{\fnm{P. } \sur{Mohanty}}\email{priyabrata120689@gmail.com}

\author*[2]{\fnm{C.} \sur{Dash}}\email{anuchinu20@gmail.com}

\author[1]{\fnm{A. } \sur{Anupam}}\email{atulanupam12@gmail.com}

\author*[1]{\fnm{B. B. } \sur{Sahu}}\email{bbsahufpy@kiit.ac.in}

\affil*[1]{\orgdiv{School of Applied Sciences}, \orgname{KIIT Deemed to be University},  
\city{Bhubaneswar}, \postcode{751024}, \state{Odisha}, \country{India}}

\affil[2]{\orgdiv{Department of physics}, \orgname{Anchalika Mahavidyalaya Gadia}, \city{Mayurbhanj}, \postcode{757023}, \state{Odisha}, \country{India}}

%\affil[3]{\orgdiv{Department}, \orgname{Organization}, \orgaddress{\street{Street}, \city{City}, \postcode{610101}, \state{State}, \country{Country}}}


\abstract{In this work, we have analyzed the structural properties of even-even $^{126-188}Nd_{60}$ isotopes. For this we have used axially deformed Relativistic Mean Field (RMF) model with PK1 and NL-SH parametrization. In structural properties, We have estimated and analyzed binding energy per nucleon (B.E./A), two neutron separation energy ($S_{2n}$), differential variation of two neutron separation energy ($dS_{2n}$), quadrupole deformation parameter ($\beta_{2}$), root mean square nuclear charge radius ($r_{ch}$), neutron skin thickness ($r_{np}$) and single particle energy (SPE) levels of Nd isotopes. Some bulk properties are also compared with experimentally accessible results and with results of Finite Range Droplet Model (FRDM). To understand the shape evolution around N = 92, the variation of the potential energy curves (PECs) with quadrupole deformation parameter are also investigated. From all the investigations, We observe some sign of stability at N = 92 and shape transitions around it.}

%%================================%%
%% Sample for structured abstract %%
%%================================%%




\maketitle
%\begin{multicols}{2}
\section{Introduction}\label{sec1}

Being in a transitional region between spherical and well-deformed nuclei, the rare earth elements (typically $Z\sim 57-64$) bear fundamental importance. Here the single-particle and collective degrees of freedom strongly interplay to govern the shape of a nucleus. The display of rich structural phenomena such as shape coexistence \cite{naz2018microscopic},
 rapid evolution of quadrupole deformation \cite{dash2024examining} 
 and the emergence of rotational behavior by these nuclei \cite{delaroche2010structure} 
  makes them the testing ground of different theoretical nuclear models and some experiments \cite{al1983levels, snelling1983gamma, ahmad1988coulomb, pitz1990low}. 
  In particular, nuclei near neutron number $N \sim 90$ are known to display signatures of quantum phase transitions \cite{casten2007quantum, ouhachi2018nuclear} 
, providing a unique opportunity to investigate critical point symmetries and shape evolution.  
For these rare earth nuclei often multiple minimas and soft deformation characteristics \cite{naz2018microscopic} 
 are observed, offering insights into fission barriers, excitation mechanisms, and nuclear stability. They also play a significant role in astrophysical processes such as r-process nucleosynthesis, as well as in applications involving nuclear energy and isotope production. Here we have chosen neodymium (Nd) in the mass number range $126 \leq A \leq 188$ as Nd isotopes with neutron number around $N \sim 90$ are identified as an example of shape transition between spherical and axially deformed nuclei \cite{ouhachi2018nuclear}. 
 
 	Many researchers also predicted the shell closure around N = 92 for Z = 62 \cite{swain2025radial}. 
  The main purpose of our present study is to investigate about the existence of N = 92 shell/sub-shell closure by examining the structural properties and shape evolution of even-even neodymium isotopes in this $^{126-188}Nd_{60}$ isotopic series. In this context we have estimated and analyzed some bulk properties such as binding energy per nucleon (B.E./A), two neutron separation energy ($S_{2n}$), differential variation of two neutron separation energy ($dS_{2n}$), nuclear charge radius($r_{ch}$), quadrupole deformation parameter ($\beta_{2}$) and neutron skin thickness ($r_{np}$). Apart from these ground state properties, the single particle energy levels of $^{142}Nd$ and $^{152}Nd$ isotopes are plotted and analyzed for zero deformation. Along with these we have also investigated the potential energy curves (PECs) of $^{140-162}Nd$ isotopes. 
 
 	There are many theoretical models which deal with nuclear structure. Among them we have considered the Relativistic Mean Field (RMF) model because it is one of the successful theoretical models in describing the ground state properties of nuclei over the periodic table \cite{gambhir1990relativistic, lalazissis1997new}. %Es paper 22, 23ref}.
Earlier it has been successfully used by our collaborators in explaining structural properties of different elements \cite{swain2018nuclear, dash2023study, dash2021structural, dash2021shell, dash2022probing, dash2024nuclear, dash2023nuclear, dash2026ground}. 
Some important features of RMF model behind its world wide acceptance are;
\begin{itemize}
\item The model considered a nucleus of being an aggregation of Dirac nucleons and they interact via the exchange of mesons in a relativistically covariant manner.
\item The fields, characterized by their angular momentum, parity and isospin mediate the interaction.
\item The model naturally includes the spin orbit interaction. 
\end{itemize}
As RMF model is a parameter dependent model, we have used two nonlinear force parameter sets such as PK1 \cite{long2004new} and NL-SH \cite{sharma1993rho}
during our calculations.

	The article is structured in such a way that section-2 describes the detailed formulation, section-3 contains the results and their analysis. Lastly the findings are concluded in section-4.

\section{Mathematical Formulation}\label{sec2}

At first the Relativistic Mean Field (RMF) model was proposed by Walecka in 1974. This model was also popularly called as $\sigma - \omega$ model \cite{walecka1974theory, serot1986adv}. 
This model neglected the self interactions among mesons which results a greater incompressibility for nuclear matter. To overcome this difficulty, Boguta and Bodmer \cite{boguta1977relativistic}
 introduced nonlinear self-coupling terms for $\sigma$-meson in the previously proposed Lagrangian density of Walecka model. 
In recent years, the nonlinear couplings of the $\omega$ mesons are also included \cite{sugahara1994relativistic}
so that it can help in reproducing the density dependence of the scalar and vector
potentials of the relativistic Brueckner-Hartree-Fock calculations \cite{brockmann1992relativistic}. 
We have started our calculations from the Lagrangian density \cite{serot1992relativistic, ring1996relativistic, patra1991relativistic, swain2018nuclear} given below;
\begin{eqnarray}
 \mathcal{L}=\bar{\psi_{i}}(i\gamma^{\mu}\partial_{\mu}-M)\psi_{i}+\frac{1}{2}
 \partial^{\mu}\sigma\partial_{\mu}\sigma-\frac{1}{2}m_{\sigma}^{2}\sigma^{2}   
 -\frac{1}{3}g_{2}\sigma^{3}-\frac{1}{4}g_{3}\sigma^{4}-g_{s}\bar\psi_{i}\psi_{i}\sigma \\  \nonumber
 -\frac{1}{4}\Omega^{\mu\nu}\Omega_{\mu\nu} 
 +\frac{1}{2}m_{\omega}^{2}V^{\mu}V_{\mu}+\frac{1}{4}C_{3}(V_{\mu}V^{\mu})^{2}
 -g_{\omega}\bar\psi_{i}\gamma^{\mu}\psi_{i}V_{\mu} 
 -\frac{1}{4}\overrightarrow{B}^{\mu\nu}.\overrightarrow{B}_{\mu\nu} \\  \nonumber
 +\frac{1}{2}m_{\rho}^{2}\overrightarrow{R}^{\mu}.\overrightarrow{R_{\mu}}
 -g_{\rho}\bar{\psi_{i}}\gamma^{\mu}\overrightarrow{\tau}\psi_{i}.\overrightarrow{R^{\mu}} 
 -\frac{1}{4}F^{\mu\nu}F_{\mu\nu}-e\bar{\psi_{i}}\gamma^{\mu}
 \frac{(1-\tau_{3i})}{2}\psi_{i}A_{\mu}  \nonumber \label{eq1}
 \end{eqnarray}
 Here the nucleons are described as Dirac spinors ($\psi$) interacting with each other via the exchange of mesons. The interaction terms $\bar{\psi}\sigma\psi$, $\bar{\psi}\gamma^{\mu}\psi$ respectively represent the coupling terms between $\sigma$-meson and $\omega$- meson with nucleons. The former gives rise to a strong attraction and the later to a strong repulsion. The $\rho$-mesons couple to the isovector current and also
with photons producing the electromagnetic interaction. It is also assumed that the nucleons move independently in these meson fields and each are considered as single particle Dirac spinors $\psi_{i}$ (i= 1, 2, 3....A) collectively forming
a slater determinant. Where A represents the number of nucleons for a particular nucleus. The arrow bars indicate the isovector quantities.
Where M, $m_{\sigma}$, $m_{\omega}$ and $m_{\rho}$ represent the masses of the nucleon, $\sigma$, $\omega$ and $\rho$ mesons respectively.
The coupling constants are represented as $g_{\sigma}$, $g_{\omega}$, $g_{\rho}$ and $\frac{e^{2}}{4\pi}$ = 1/137 for $\sigma$, $\omega$, $\rho$ mesons and for the photon respectively. 
The field for $\sigma$, $\omega$ and $\rho$ mesons are respectively represented by $\sigma$, $V^{\mu}$ and $\vec{R^{\mu}}$.
The electromagnetic field ($A^{\mu}$)couples to the protons.   

$\Omega^{\mu\nu}$, $\overrightarrow{B}^{\mu\nu}$ and $F^{\mu\nu}$ respectively represent the field tensors of $\omega$-meson, $\rho$-meson and electromagnetic field.
These are represented below;
\begin{equation}
\Omega^{\mu\nu}=\partial^{\mu}V^{\nu}-\partial^{\nu}V^{\mu} \label{eq2}
\end{equation}
\begin{equation}
\overrightarrow{B}^{\mu\nu}=\partial^{\mu}\overrightarrow{R}^{\nu}-\partial^{\nu}\overrightarrow{R}^{\mu}-g_{\rho}(\overrightarrow{R}^{\mu}\times \overrightarrow{R}^{\nu}) \label{eq3}
\end{equation}
\begin{equation}
F^{\mu\nu}=\partial^{\mu}A^{\nu}-\partial^{\nu}A^{\mu} \label{eq4}
\end{equation}
The application of classical variational principle to the above Lagrangian density yields the Klein Gordan equation and Dirac equation for mesons and nucleon respectively. The contribution of antiparticles are neglected during the calculations. The static solutions of these field equations provides us the ground state observables of a nucleus. Considering the Dirac spinors as eigen vectors, the single particle energies are obtained as eigen values of static Dirac equation. Here, the wave functions are expanded in a deformed harmonic oscillator basis taking maximum oscillator shells for both bosons and fermions as 14. The solutions are carried out by a self consistent iteration method with initial deformation value \cite{serot1986adv, boguta1977relativistic, del2001pairing} 
$\beta_{0}$ = -0.3, -0.2, -0.1, 0.0, 0.1, 0.2, 0.3. From these solutions, we obtain different physical properties. 

	The physical quantity assigned to measure the departure of nuclear charge distribution from the spherical symmetry is the quadrupole moment (Q). The deformation parameter $\beta$ is related to Q with the equations mentioned below;

\begin{equation}
Q=Q_{n}+Q_{p}=\sqrt{\frac{16\pi}{5}}\big(\frac{3}{4\pi}AR^{2}\beta\big) \label{eq5}
\end{equation}
where R is known as radius of the nucleus and it is $R=1.2A^{\frac{1}{3}}$. 
The proton and neutron deformation parameters can be expressed as follows 
\begin{equation}
\beta_{p}=\sqrt{5\pi}\frac{Q_{p}}{3ZR^{2}} \label{eq6}
\end{equation}
\begin{equation}
\beta_{n}=\sqrt{5\pi}\frac{Q_{n}}{3NR^{2}} \label{eq7}
\end{equation}
Integrating over the distance of proton and neutron density distribution, we estimate respectively the root mean square (rms) proton and neutron radius. The charge radius is given as;
\begin{equation}
r_{ch}=\sqrt{r_{p}^{2}+0.64} \label{eq8}
\end{equation} 
where $r_{p}$ is proton radius and it is expressed by the equation below
\begin{equation}
\langle r_{p}^{2}\rangle=\frac{1}{Z}\int{\rho_{p}(r_{\perp},z)r_{p}^{2}d\tau_{p}} \label{eq9}
\end{equation} 
similarly we can express neutron radius ($r_{n}$)
\begin{equation}
\langle r_{n}^{2}\rangle=\frac{1}{N}\int{\rho_{n}(r_{\perp},z)r_{n}^{2}d\tau_{n}} \label{eq10}
\end{equation}

	The total energy of the system is;
\begin{equation}
E_{total}=E_{part}+E_{\sigma}+E_{\omega}+E_{\rho}+E_{C}+E_{pair}+E_{c.m} \label{eq11}
\end{equation}
Where 
\begin{equation}
E_{part}=\sum\limits_{i}\textit{v}_{i}^{2}\int{d^{3}r\psi_{i}^{\dagger}\big[-\vec{\alpha}.\vec{\nabla}-\vec{\alpha}.\vec{V}+\beta M^{*}+V(r_{\perp} ,z)\big]\psi_{i}} \label{eq12}
\end{equation}
\begin{equation}
E_{\sigma}=\int{d^{3}r}\big[\frac{1}{2}(\nabla\sigma)^{2}+U(\sigma)\big] \label{eq13}
\end{equation}
\begin{equation}
E_{\omega}=-\int{d^{3}r\big[\frac{1}{2}(\nabla V_{0})^{2}+\frac{1}{2}m_{\omega}^{2}V_{0}^{2}+C_{3}V_{0}\big]} \label{eq14}
\end{equation}
\begin{equation}
E_{\rho}=-\int{d^{3}r\big[\frac{1}{2}(\nabla\rho_{0})^{2}+\frac{1}{2}m_{\rho}^{2}\rho_{0}^{2}\big]} \label{eq15}
\end{equation}
\begin{equation}
E_{C}=-\int{d^{3}r\big[\frac{1}{2}(\nabla A_{0})^{2}\big]} \label{eq16}
\end{equation}
\begin{equation}
E_{pair}=-G\bigg[\sum\limits_{i>0}\textit{u}_{i}\textit{v}_{i}\bigg]^{2} \label{eq17}
\end{equation}
and
\begin{equation}
E_{c.m}=-\frac{3}{4}41A^{\frac{-1}{3}} \label{eq18}
\end{equation}
Where $E_{part}$ represents the energy of the nucleons moving in meson fields.

$ E_{C}, E_{\sigma}, E_{\omega}, E_{\rho},$ represents the energies of the Coulomb field and corresponding meson fields.
$E_{pair}$ represents pairing energy with pairing force of strength G. $\textit{v}_{i}^{2}$ and $\textit{u}_{i}^{2}=1-\textit{v}_{i}^{2}$ represents the occupation and nonoccupational probabilities.
To obtain the numerical solutions the source code used is from Ref \cite{ring1997computer}. 

\section{Results and Discussion}\label{sec3}
In this work, the bulk properties such as binding energy per nucleon (B.E./A), neutron skin thickness ($r_{np}$), nuclear charge radius ($r_{ch}$), quadrupole deformation parameter ($\beta_{2}$), two neutron separation energy ($S_{2n}$) and differential variation of two neutron separation energy ($dS_{2n}$) have been estimated and analyzed for even-even $^{126-188}Nd_{60}$ isotopes. The calculations are made with two non linear force parameter sets that are PK1 \cite{long2004new} and NL-SH \cite{sharma1993rho}.
 The single particle energies for $^{142}Nd$ and $^{152}Nd$ have been investigated for zero deformation to have a conclusive evidence of shell closure. Apart from that we have also studied about the shape evolution of Nd isotopes around the transition region near $N \sim 90$ by plotting PECs with quadrupole deformation parameter for $^{140-162}Nd$ isotopes. 

\subsection{Binding Energy}\label{subsec2}
\begin{figure}[h!]
\centering
\includegraphics[width=0.9\textwidth, angle=-90]{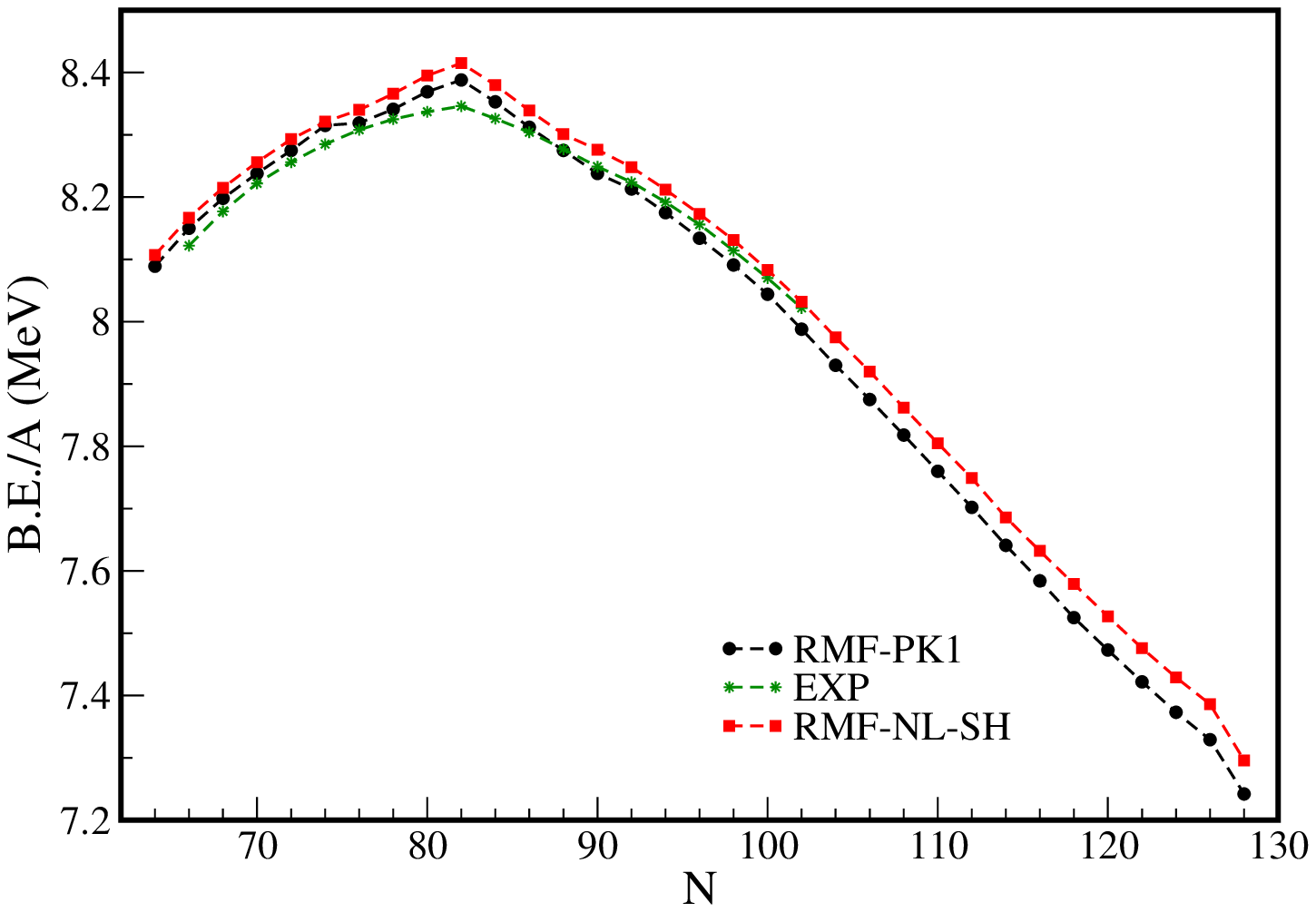}
\caption{Variation of B. E./A with neutron number of Nd, estimated for RMF model with PK1 and NL-SH parameter set and compared with experimental values obtained from National Nuclear Data Centre (NNDC) \cite{nndc}.}\label{fig1}
\end{figure}
Binding energy is one of the important and fundamental property of a nucleus. Being among one of the measurable quantities, B.E. establishes the validity of theoretical models. Here in Fig. \ref{fig1}, the variation of B.E./A shows a small peak at N = 82, the pronounced neutron magic number. The broken non-linearity for PK1 parameter set is also observed at N = 74 and 126. From the figure it is well observed that our estimated results are in good agreement with the available experimental results obtained from National Nuclear Data Centre (NNDC) \cite{nndc}. 
For PK1 and NL-SH the maximum error, we found is  0.042 MeV and 0.069 MeV respectively.

\subsection{Two Neutron Separation Energy} \label{subsec3}
Two neutron separation energy ($S_{2n}$) is an essential parameter towards the structural understanding of a nucleus especially for those lying away from the stability line. It provides us the idea regarding the energy gap between the adjacent shells and a large shell gap is a sign of shell/sub-shell closure. As $S_{2n}$ is free from odd even staggering (OES), it provides better platform to locate neutron drip line, shell closures and to search for shape transitions because many a times they appear as kinks in the variation of $S_{2n}$ with neutron number. Substituting the estimated B.E. values in the equation below, we have calculated $S_{2n}$ for both parameter sets.
\begin{equation}
S_{2n}=B.E.(N,Z)-B.E.(N-2,Z) \label{eq19}
\end{equation}
\begin{figure}[h!]
\centering
\includegraphics[width=0.9\textwidth, angle=-90]{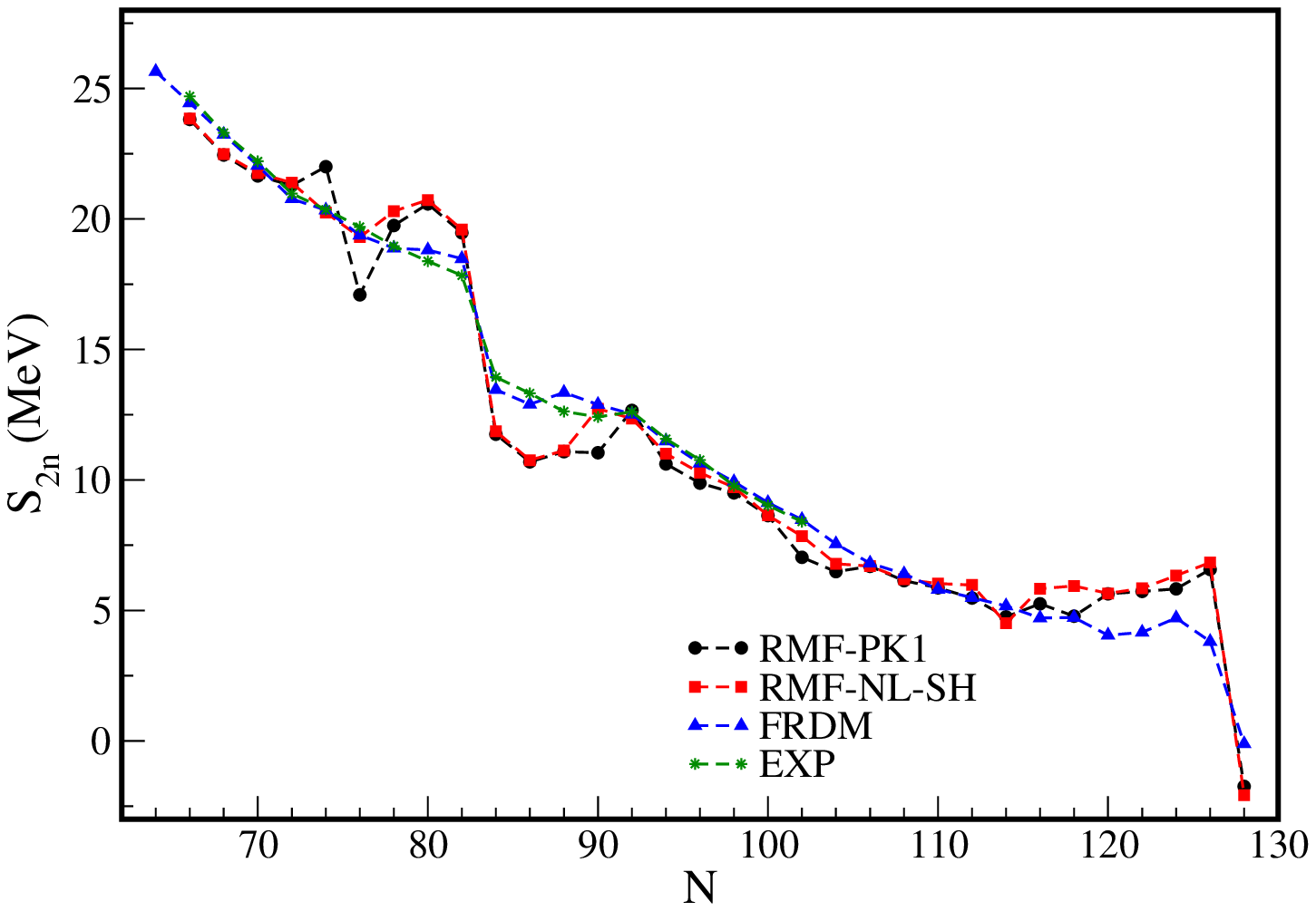}
\caption{Variation of $S_{2n}$ with neutron number of Nd, estimated for RMF model with PK1 and NL-SH parameter set and compared with experimental values obtained from National Nuclear Data Centre (NNDC) \cite{nndc} and with that obtained for FRDM \cite{moller2016nuclear}.}\label{fig2}
\end{figure}
In Fig. \ref{fig2}, two large drops at N = 82 and N = 126 corresponding to two neutron magic numbers are clearly observed for all four curves. Except these, we observe another sudden fall at N = 74 for PK1 parameter. But it does not manifest itself as an energy gap in the single particle energy plot in Fig. \ref{fig4}. 
So, we thought that it may be associated with some structural changes \cite{ouhachi2018nuclear}  
around the isotope. However a deep at N = 76 for both PK1 and NL-SH curves can be correlated with shape transition from prolate at N = 76 to oblate at N = 78 shown in Fig. \ref{fig7}.
Apart from that we observe a deep at N = 92 for both parameter sets. We can relate this  deep with the gap in single particle energy level indicating possible shell/ sub-shell closure. Previously the shell/sub-shell closure at N = 92 is observed for Sm isotope by our collaborators \cite{swain2025radial}. 
To get further insight into the shell structure and shape transition we have estimated and analyzed the differential variation of two neutron separation energy ($dS_{2n}$). Large drops of $S_{2n}$ manifest as large deeps in $dS_{2n}$
providing confirmation of the results obtained from the variation of $S_{2n}$.
The differential variation of two neutron separation energy ($dS_{2n}$) is calculated using $S_{2n}$ values in the equation below;
\begin{equation}
dS_{2n}=\frac{S_{2n}(N+2,Z)-S_{2n}(N,Z)}{2} \label{eq20}
\end{equation}
The variation of $dS_{2n}$ with neutron number for the isotopic series of Nd is shown in Fig. \ref{fig3}.
\begin{figure}[h!]
\centering
\includegraphics[width=0.9\textwidth, angle=-90]{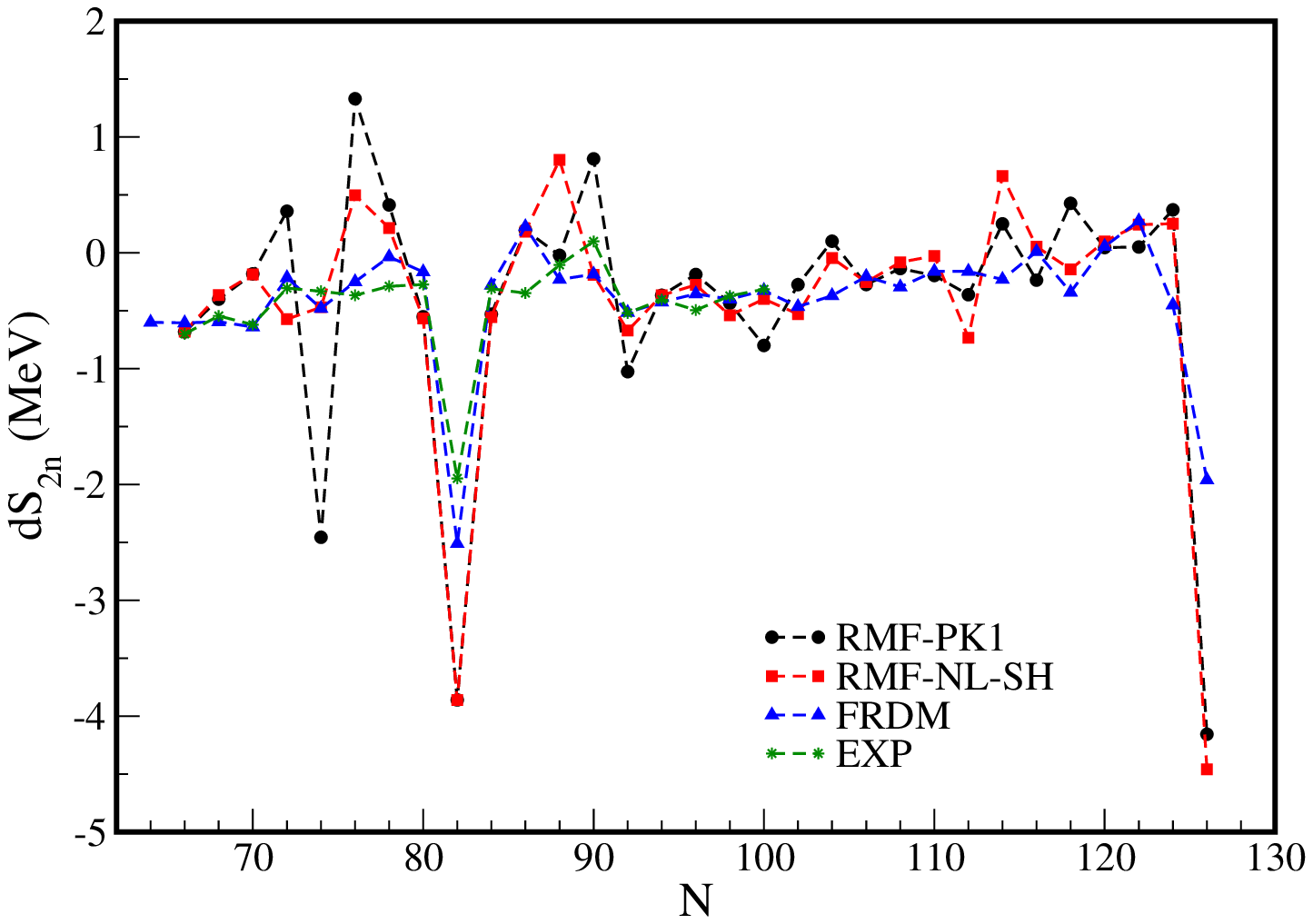}
\caption{Variation of $dS_{2n}$ with neutron number of Nd, estimated for RMF model with PK1 and NL-SH parameter set and compared with experimental values obtained from National Nuclear Data Centre (NNDC) and with that obtained for FRDM \cite{moller2016nuclear}.}\label{fig3}
\end{figure}
A small but clear deep at N = 92 is observed for all four curves except a pronounced deep at N = 82, the neutron magic number. Some times this irregularity around N = 92 is connected with structural changes around these isotopes \cite{ouhachi2018nuclear}. 
From several studies also, $N \sim 90$ has been emerged as an example of the critical point X(5) phase/shape transitions \cite{krucken2002b, fossion20065, nikvsic2007microscopic, robledo2008evolution, rodriguez2008beyond}. 
In addition to the above observations, PK1 curve shows kinks at N = 74, 100, 112, 116 and NL-SH curve at N = 72, 98, 112 and 118.
\subsection{Single Particle Energy}\label{subsec4}
\begin{figure*}[h!]
\centering
\begin{tabular}{cc}
\includegraphics[width=0.5\textwidth, height=7cm]{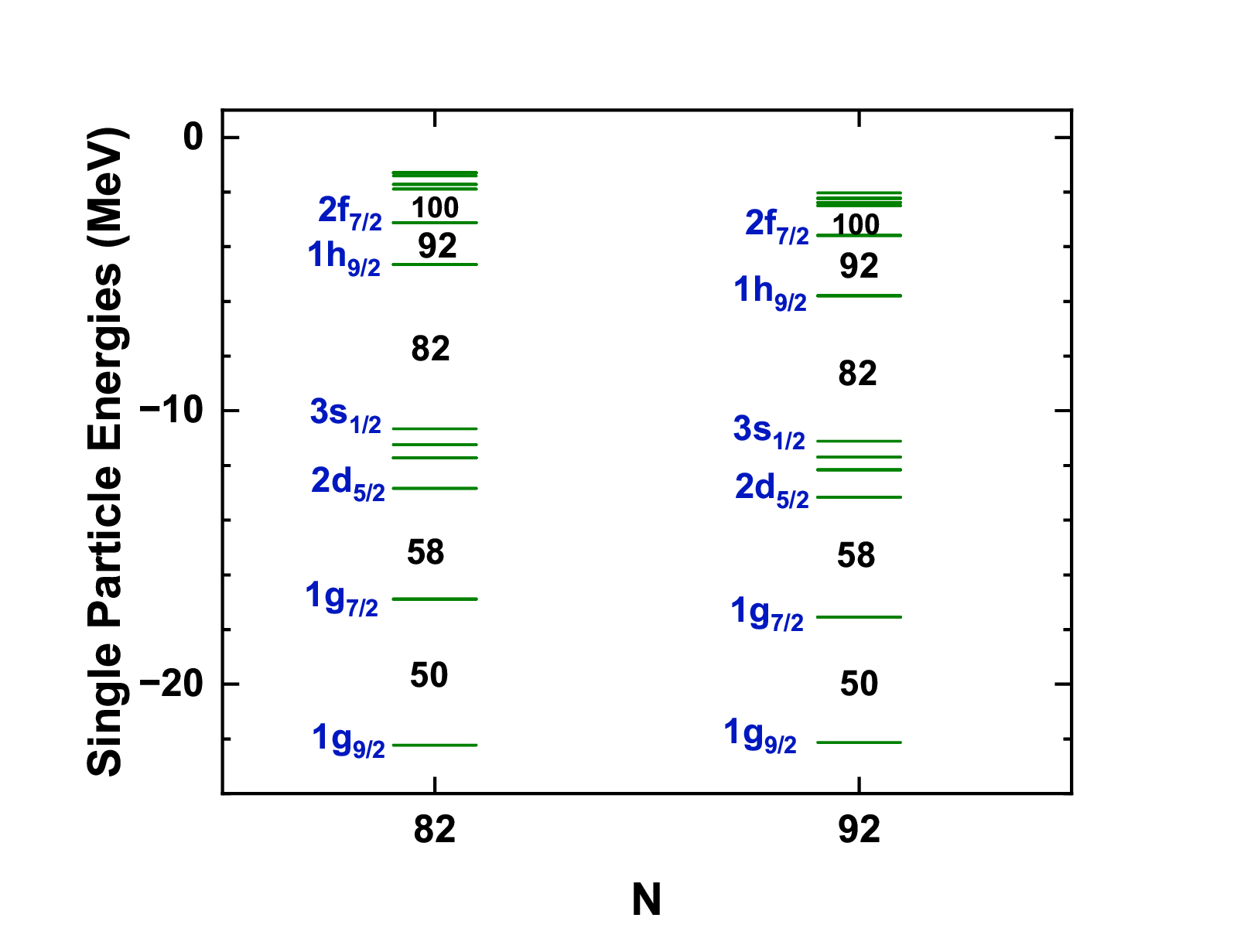} & 
\includegraphics[width=0.5\textwidth, height=7cm]{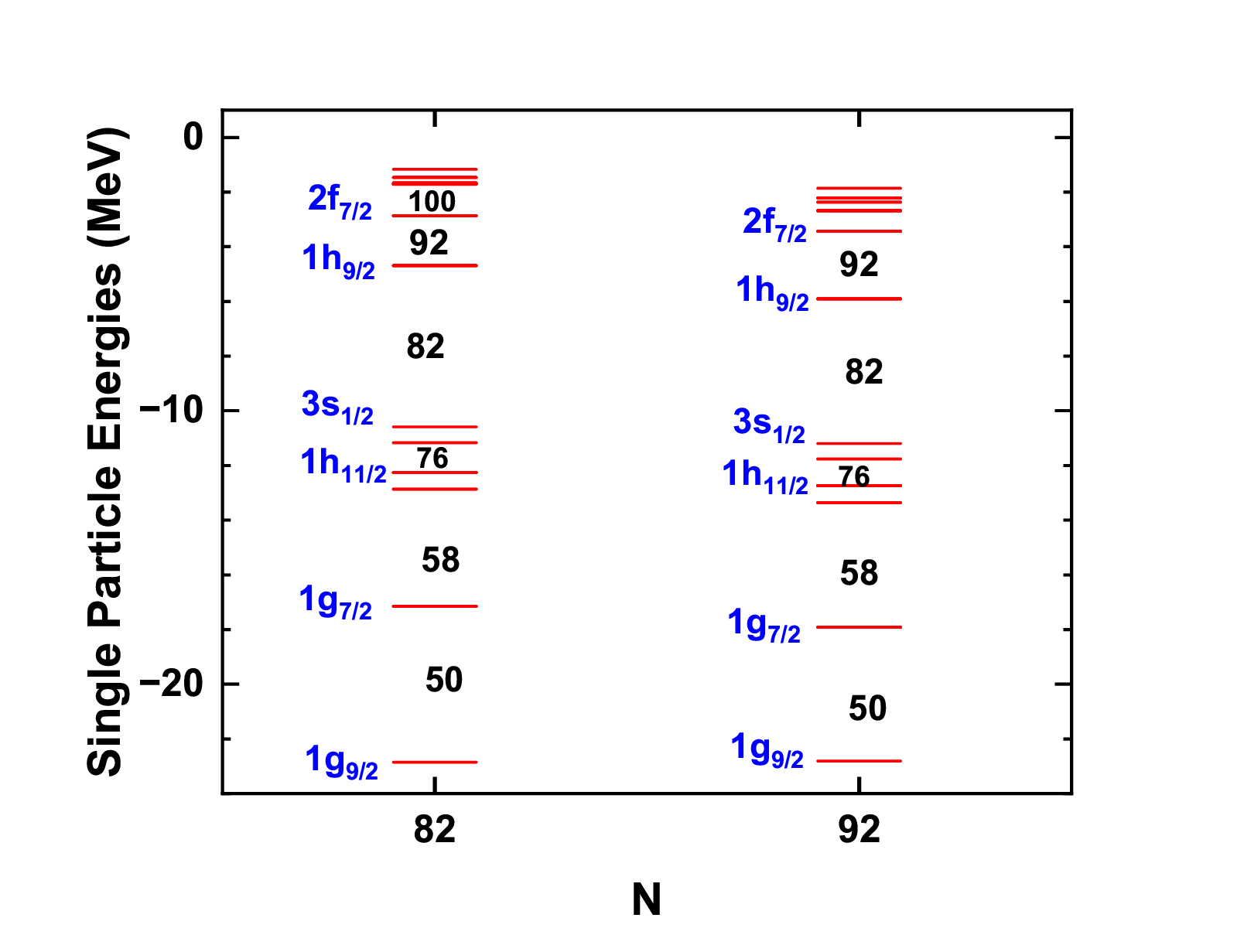} \\ 
(a) PK1  & (b) NL-SH \\
\end{tabular}
\caption{Single particle energy levels of $^{142}Nd$ and $^{152}Nd$}\label{fig4}
\end{figure*}

To get more insights into the structural stability of the isotopic series, the neutron single-particle energy levels of $^{142}Nd$ and $^{152}Nd$ were plotted in Fig. \ref{fig4}(a) and Fig. \ref{fig4}(b) for PK1 and NL-SH parameter set respectively. 

The analysis focuses on four key aspects: direct energy level comparisons, systematic shifts induced by the addition of  neutrons, the sensitivity of orbital ordering to the choice of parameter set, and the resulting implications for the understanding of shell structure and level spacing in this region of the nuclear chart.

	The doubly-magic-like nucleus $^{142}Nd$, with its closed neutron shell at N = 82, provides a foundational reference point for analyzing the mean-field potential deep within a major shell closure \cite{cocolios2011early}. 
The analysis reveals that while the overall structure of the deeply bound orbitals is similar, significant differences emerge in the mid-shell region, which governs the properties of nearby nuclei. 
For the lowest-lying orbitals, the energy differences between the PK1 and NL-SH predictions are relatively small. 
The splitting between the $1g_{9/2}$ and $1g_{7/2}$ orbitals, critical for understanding the Z = 60 proton shell, is approximately 5 MeV for PK1 and 6 MeV for NL-SH. 
This close agreement suggests that the bulk properties of the mean field, which primarily determine the energies of these deeply bound states, are well-constrained by both parameter sets. 
Similarly, the ordering and separation of the $1h_{11/2}$ and $1h_{9/2}$ orbitals remain consistent, with both functionals predicting the $1h_{11/2}$ state to be significantly lower in energy than the $1h_{9/2}$ state. 

These small discrepancies indicate a robust prediction for the low-lying part of the spectrum, reflecting a shared foundation in the underlying physics of the RMF approach. 

	However, moving towards the higher-energy mid-shell orbitals, the divergence between the two parameter sets becomes more pronounced, highlighting a region of greater model dependence.

The transition from $^{142}Nd$ (N = 82) to $^{152}Nd$ (N = 92) represents the filling of a new major sub-shell, driven by the formation of a sub-shell closure at N=92 . This process induces significant changes in the s.p. spectrum, altering orbital energies and their relative order. 
Across nearly all labeled orbitals, the addition of neutrons from N = 82 to N = 92 causes a substantial downward shift in s.p. energies for both the PK1 and NL-SH parameter sets. 
This is a direct consequence of increased neutron density, which enhances the attractive mean field and lowers the overall binding energy. 
The shift is most pronounced for orbitals that were previously unoccupied or high-lying in $^{142}Nd$. 
For example, the $1h_{11/2}$ orbital, which lies far above the Fermi surface in $^{142}Nd$, drops by over 2.0 MeV in energy when transitioning to $^{152}Nd$ according to both models. 
This dramatic lowering indicates a strong interaction between the newly added valence neutrons and this particular orbital, stabilizing it significantly within the new potential landscape. 
Similar large downward shifts are observed for other high-lying orbitals, reinforcing the conclusion that the primary driver of spectral evolution in this mass region is the progressive filling of the N = 92 sub-shell \cite{Zhang2020TowardAN}. 

	The NL-SH parameter set predicts a marked reordering of mid-shell orbitals in $^{152}Nd$. 
 
Specifically, the $1h_{9/2}$ orbital is pushed down relative to the $1h_{11/2}$ orbital, further modifying the s.p. structure near the new Fermi surface. 
This suggests that the NL-SH model favors a configuration where the f-orbitals play a more prominent role in defining the N = 92 shell stabilization. 
This prediction aligns with studies of shape evolution in neighboring samarium isotopes, where examining the single-particle spectra is key to understanding structural changes \cite{meng2003first}. 
In contrast, the PK1 parameter set predicts a more gradual evolutionary path. 
While the general trend of lowering orbital energies holds true, the relative ordering of the mid-shell orbitals in $^{152}Nd$ remains more similar to that observed in $^{142}Nd$ compared to the predictions of NL-SH. 
For instance, the $2f_{7/2}$ orbital, while lowered, does not drop as precipitously as it does in the NL-SH calculation, maintaining a larger energy gap above the $3s_{1/2}$ orbital. 
This implies that the mean-field potential generated by PK1 changes more smoothly as the shell fills, without inducing the same sharp rearrangement of orbitals seen with NL-SH. 
This difference in evolutionary trajectory highlights a fundamental ambiguity in modeling the precise nature of the N = 92 sub-shell closure; both functionals are consistent with the general phenomenon of shell filling, but they differ in the specific composition and stability of the orbitals that define this new shell.

This data shows that the N = 82 gap is larger under the NL-SH prediction than that of PK1. 
For the N = 92 case, the gap predicted by NL-SH is again larger than that predicted by PK1 , reinforcing the idea that NL-SH favors a more pronounced shell closure. 
The fact that the N = 92 gap is smaller than the N = 82 gap in both models is expected, as the N = 92 "magic" number is deformed and thus less stable than the spherical N = 82 shell. 
However, the model-dependent nature of these gaps highlights a significant uncertainty in the precise location and strength of this new shell effect. 
Theoretical studies have explored the properties of new magic numbers, and the N = 92 pseudo-shell is a topic of ongoing research in the context of exotic nuclei \cite{zhang2016global}.  
The disagreement between PK1 and NL-SH exemplifies the challenges in pinning down these properties with current theoretical tools.
\subsection{Nuclear Charge Radius and Neutron Skin Thickness}\label{subsec5}
\begin{figure}[h!]
\centering
\includegraphics[width=0.9\textwidth, angle=-90]{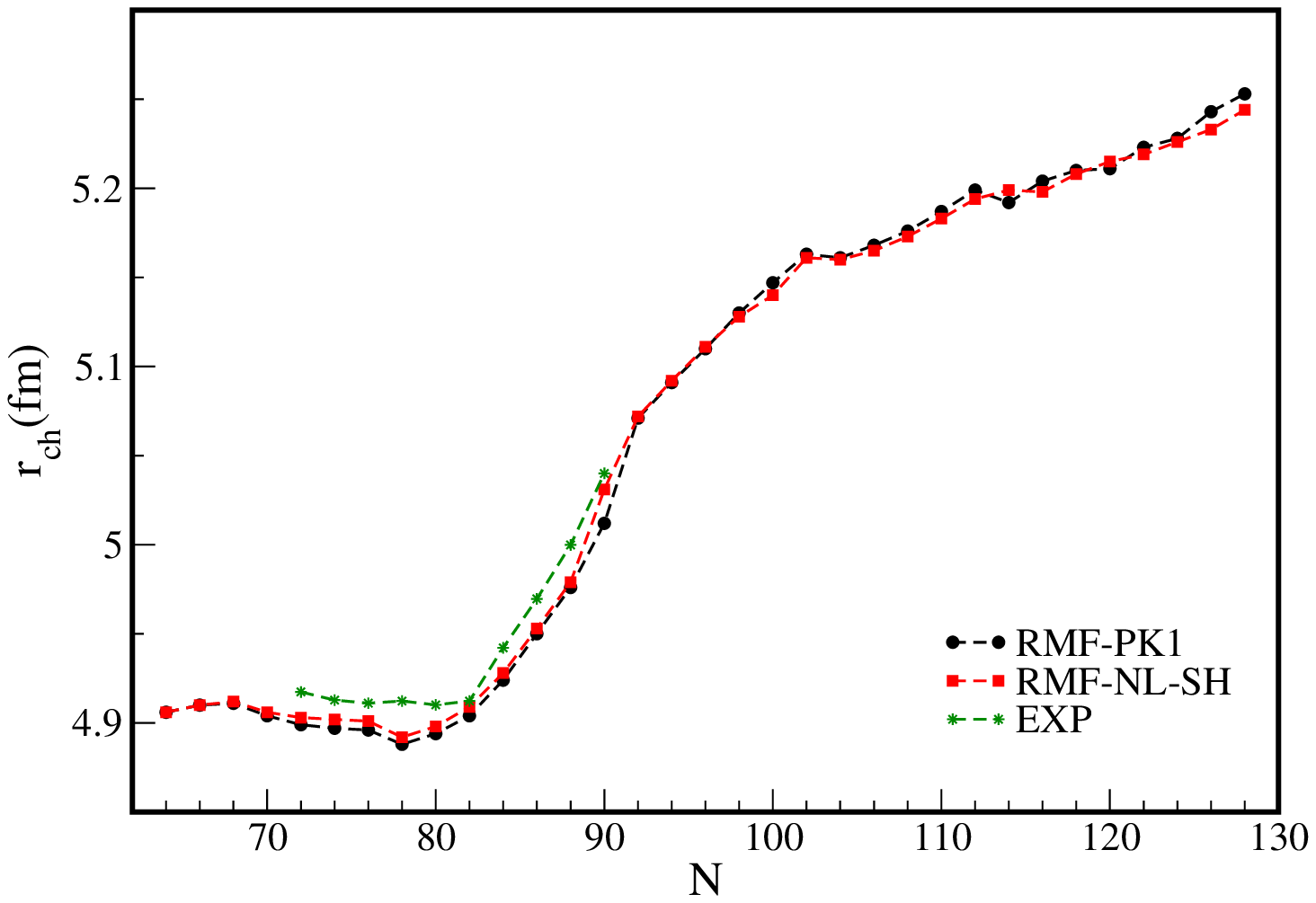}
\caption{Variation of $r_{ch}$ with neutron number of Nd, estimated for RMF model with PK1 and NL-SH parameter set and compared with experimental values obtained from \cite{angeli2013table} }\label{fig5}
\end{figure}

The rms radius is a sensitive tool towards the change of shape and size of an isotope in an isotopic series \cite{rodriguez2010charge}. 
We have thus estimated the charge radius ($r_{ch}$).
Nuclear charge radius ($r_{ch}$) is also sensitive towards the shell effect in an isotopic series. In some cases, $r_{ch}$ is found to manifest itself as a kink across
spherical shell closures \cite{kreim2014nuclear, gorges2019laser, reponen2021evidence, day2021laser}. 
Around N = 40 sub-shell closure, $r_{ch}$ shows a localized effect for Nickel isotope in laser spectroscopy measurements by Jessica Warbinek et al \cite{warbinek2024smooth} %{Es paper4]}
relative to droplet model \cite{malbrunot2022nuclear, yang2023laser}. 
 While at N = 32 sub-shell closure, in neuron rich Potassium isotope \cite{koszorus2021charge} 
the charge radius doesn't manifest itself. In order to check the manifestation of $r_{ch}$ around N = 92 which is reflected in separation energies and single particle energy levels, we have plotted $r_{ch}$ with neutron number in Fig. \ref{fig5}. 
Our estimated results are found to be in good qualitative agreement with the experimentally accessible results \cite{angeli2013table}. It is observed that all the three curves show large slope from N = 82 to N = 92 indicating a rapid change in size of charge density distribution of the nucleus with increasing neutron count. Apart from that we observe two small kinks at N = 102 and N = 112 for both parameters. Very small but variations in the slope of the curves are observed around N = 92 indicating a shell/sub-shell closure.

\begin{figure}[h!]
\centering
\includegraphics[width=0.9\textwidth]{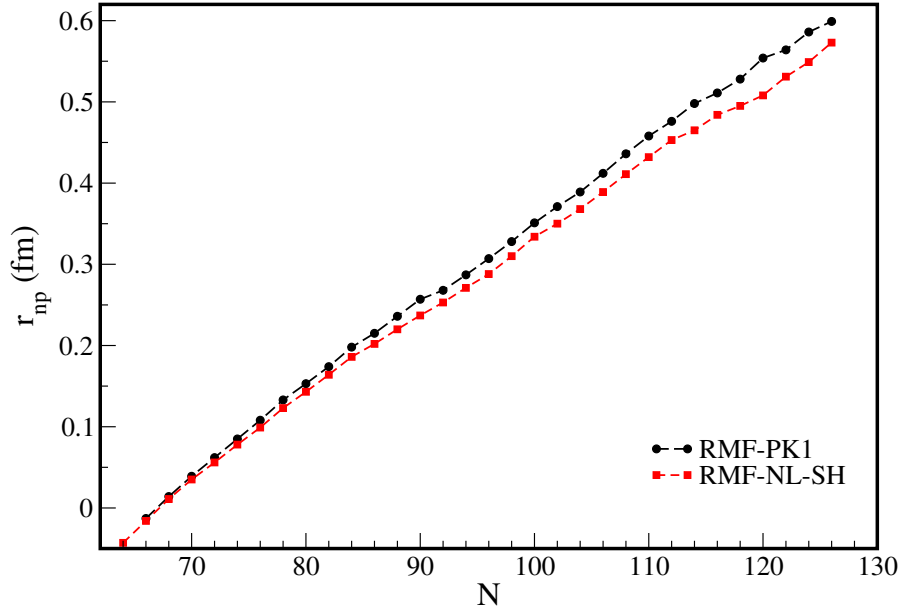}
\caption{Variation of $r_{np}$ with neutron number of Nd, estimated for RMF model with PK1 and NL-SH parameter set.}\label{fig6}
\end{figure}

%%%%%%%%%%%%%%%%%%%%%%5
The neutron skin thickness ($r_{np}$) can be defined as the difference between root mean square (rms) nuclear radii estimated using the density distributions for point neutrons and point protons and can be calculated following the equation below;
\begin{equation}
r_{np} = r_{n}-r_{p} \label{eq21}
\end{equation} 
Where $r_n$ is neutron rms radius and $r_p$ is proton rms radius. In case of neutron rich nuclei, the large neutron excess comes out as neutron skin where the neutron density is expected to extend beyond the proton density \cite{reinhard2016nuclear, hagen2015charge}. 
The variation of neutron skin thickness with neutron number of Nd is shown in Fig. \ref{fig6}. $r_{np}$ increases linearly with increasing neutron number indicating the pressure of symmetry energy. Besides a very small but broken linearity around $N \sim 92$, small kinks at N = 120 and 124 are observed for PK1 parameter set. For NL-SH it is around N = 120. 
\subsection{Quadrupole Deformation Parameter}\label{subsec6}
\begin{figure}[h!]
\centering
\includegraphics[width=0.9\textwidth, angle=-90]{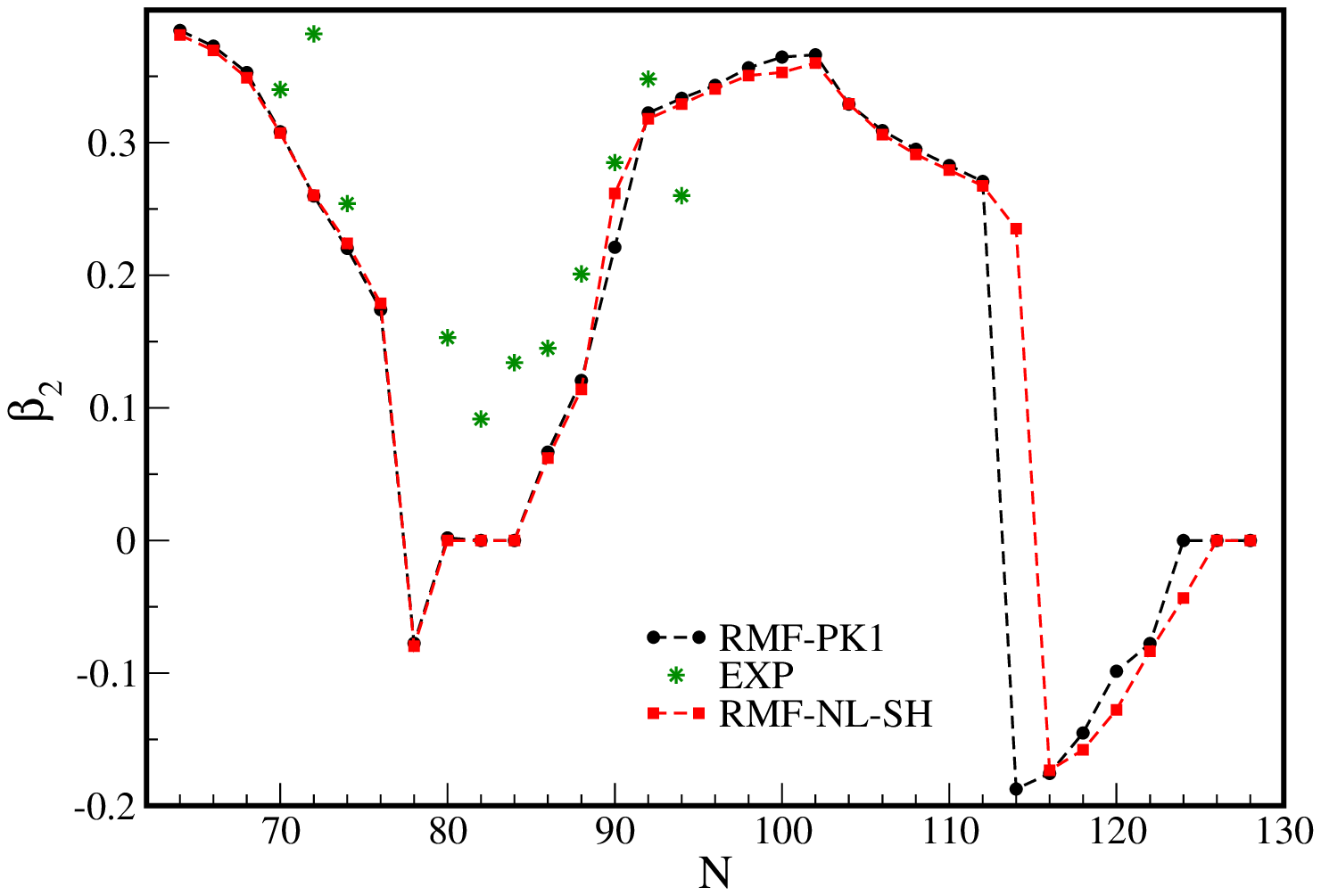}
\caption{Variation of $\beta_{2}$ with neutron number of Nd, estimated for RMF model with PK1 and NL-SH parameter set.}\label{fig7}
\end{figure}

Along with mass and charge radius, the shape is one of the prime properties of a nucleus. It is the result of collective interplay of macroscopic properties and microscopic properties like shell effect. One of the  essential characteristics to understand shape transitions is quadrupole deformation parameter. Nuclear shape evolutions along the neodymium isotopic chain is illustrated in Fig. \ref{fig7}, where the calculated quadrupole deformation parameter ($\beta_{2}$) are plotted against neutron number for both parametrization and also compared with experimentally accessible values. Here we observe shape transitions from prolate at N = 76 to oblate at N = 78 to spherical at N = 80 for both PK1 and NL-SH. Along with N = 80, the spherical shape is obtained for  N = 82 and 84. The isotopes with $N \geq 86$ acquired the prolate shape again but with a steep slope for $\beta_{2}$ with increasing neuron number upto N = 92. The prediction of rapid shape transitions suggests the occurrence of shape coexistence in the transitional region \cite{gorgen2009study}. 
To account for the configuration mixing, correlation beyond the mean field is required in the calculations \cite{gorgen2009study}. 
 From N = 92 to N = 112 for PK1 and N = 92 to N = 114 for NL-SH, the $\beta_{2}$ values are nearly equal with prolate  shape. Again the shape transition we observe from prolate at N = 112 to oblate at N = 114 then to spherical at N = 124, 126 and 128 for PK1 but for NL-SH the transition takes place from prolate at N = 114 to oblate at N = 116 then to spherical at N = 126, 128. The rapid increase of $\beta_{2}$ from N = 84 to N = 92 can be correlated with the similar behavior of charge radius. 
%\newpage
\subsection{Potential Energy Curves}\label{subsec7}
 \begin{figure*}[htbp]
\centering
\begin{tabular}{c}
\makecell{\includegraphics[width=0.85\textwidth]{shapeNdf.eps}\\ (a) $^{140-150}Nd$} \\
\makecell{\includegraphics[width=0.8\textwidth, angle=-90]{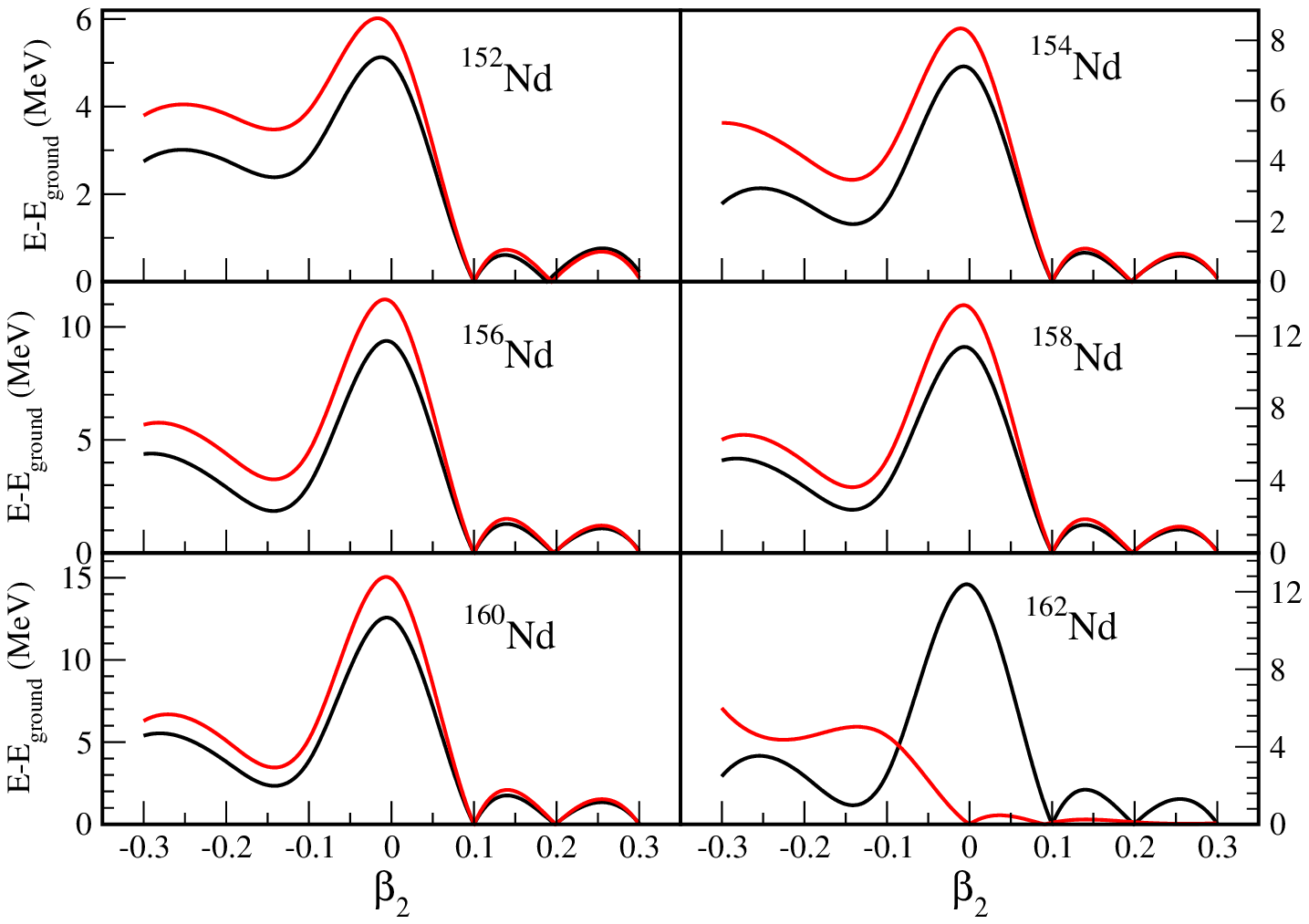}\\ (b) $^{152-162}Nd$}\\
\end{tabular}
\caption{The potential energy curves of Nd isotopes estimated for RMF model with PK1 parameter set (The solid black line) and NL-SH parameter set (The red solid line)}\label{fig8}
\end{figure*}
To understand the nuclear shape evolution around N = 92, we have plotted the potential energy curves (PECs) against quadrupole deformation parameter ($\beta_{2}$) for even even $^{140-162}Nd$ isotopes in Fig. \ref{fig8} for both parametrization. In all PECs the ground state energy is taken as a reference. From Fig. \ref{fig8} it is observed that the neodymium isotopes gradually evolve from spherical shape at N = 80 to strong prolate deformed shape at N = 92. The variation is similar for both PK1 and NL-SH parametrization. A noticeable spherical minimum is observed at N = 82 can be correlated with the shell closure which is also explained in section \ref{subsec3} and \ref{subsec4}. 
At N = 80 ($^{140}Nd$), we observe a flat base in the interval $-0.1 < \beta_{2} \leq 0.1$ for PK1 which indicates soft behavior of the isotope towards the shape transition near the equilibrium shape i.e an indication of shape co-existence at N = 80 but NL-SH curve shows a rigid spherical minimum at N = 80 i.e no sign for shape co-existence. 

One more contradictory result we get, is for $^{162}Nd$ where we observe a flat minimum lies between $0 \leq \beta{2} \leq 0.3$ for NL-SH but PK1 does not display such variation. This contradiction between two parametrization reflects the parameter dependence nature of RMF model.
From $^{146-160}Nd$ we observe minima both for prolate and oblate shapes with a spherical barrier between them. The prolate shape contains the deeper minimum indicating the ground state. For $^{146}Nd$ the energy gap between two minimas is around 0.01 MeV for NL-SH and 0.1 MeV for PK1 which reflects a strong competition between two configurations indicating a possibility of shape co-existence. Also for $^{148}Nd$ the energy gap is below 1 MeV. After that the energy gap between two minimas continues to increase with addition of neutrons. 
%\newpage
At N = 92 the gap between prolate minimum and oblate minimum is around 2.5 Mev for PK1 and 3.5 MeV for NL-SH having the deeper minimum in prolate shape. These observations establishes a strong dominance of prolate shape in Nd isotopes except some possible cases of shape co-existence. The kink from $S_{2n} \sim N$ in Fig. \ref{fig2} and $dS_{2n} \sim N$ in Fig.\ref{fig3} and an energy gap in single particle energy level in Fig. \ref{fig4} at N = 92 doesn't yield itself as a spherical minimum in PECs rather we observe a spherical barrier at zero deformation. Hence we thought that the origin of this stability is from deformed shell effects rather than spherical magicity. This picture can be made more clear from the study of variation of single particle energies with deformation parameter i.e the Nilson plot.

The shallow competing minima in axial potential energy curves may be an indication of possible $\gamma$ softness and the presence of triaxial correlations \cite{rahaman2026shape}. 
 In our present work, several isotopes are found to exhibit prolate and oblate minima separated around 1 MeV, suggesting a soft energy surface with respect to the triaxial degree of freedom. We can get a detailed insight of these features by analyzing the two dimensional ($\beta , \gamma$) plots. But for that we need a substantially different computational framework and these are beyond the scope of our present study.
\newpage
\section{Conclusion}\label{sec5}
In this article, the nuclear structural properties for even-even $^{126-188}Nd_{60}$ isotopes have been estimated using PK1 and NL-SH parametrization in axially deformed RMF model. Our results show qualitative agreement with experimentally accessible results while compared. Large drops at N = 82 and 126 in $S_{2n} \sim N$, deep in $dS_{2n}\sim N$, a large energy gap in SPE level at N = 82, zero deformation in $\beta_{2} \sim N$ and a spherical minimum for N = 82 in the PECs for both PK1 and NL-SH reproduced the pronounced spherical shell closure establishing the applicability of RMF model. Apart from this,at N = 92, a noticeable kink is observed in $S_{2n}$ and $dS_{2n}$ as function of neutron number, accompanied by a significant energy gap in SPE level. However a substantial quadrupole deformation is obtained with $\beta_{2} = 0.3224$ for PK1 and $\beta_{2} = 0.3179$ for NL-SH. Furthermore the PECs exhibit a spherical barrier of about 6 MeV in PECs at N = 92 . These features may be an indication of the presence of a deformed shell closure at N = 92. This needs further investigation. In particular a detailed analysis of evolution of single particle energies with deformation is expected to provide deeper insights into the conclusive physics behind the behavior of nuclear observables at N = 92. 
%\bibliographystyle{unsrt}
%\newpage
\bibliography{sn-bibliography}% common bib file

@article{naz2018microscopic,
  title={Microscopic description of structural evolution in pd, xe, ba, nd, sm, gd and dy isotopes},
  author={Naz, Tabassum and Bhat, GH and Jehangir, S and Ahmad, Shakeb and Sheikh, JA},
  journal={Nuclear Physics A},
  volume={979},
  pages={1--20},
  year={2018},
  publisher={Elsevier}
}

@article{dash2024examining,
  title={Examining the nuclear structure and decay modes of 128--208Sm Isotopes},
  author={Dash, C and Tripathy, G and Mohanty, P and Anupam, A and Naik, I and Sahu, BB},
  journal={International Journal of Modern Physics E},
  volume={33},
  number={12},
  pages={2450061},
  year={2024},
  publisher={World Scientific}
}

@article{delaroche2010structure,
  title={Structure of even-even nuclei using a mapped collective Hamiltonian and the D1S Gogny interaction},
  author={Delaroche, J-P and Girod, M and Libert, J and Goutte, H and Hilaire, S and P{\'e}ru, S and Pillet, N and Bertsch, GF},
  journal={Physical Review C—Nuclear Physics},
  volume={81},
  number={1},
  pages={014303},
  year={2010},
  publisher={APS}
}

@article{al1983levels,
  title={Levels in 60146Nd populated by the (n, n'$\gamma$) reaction},
  author={Al-Janabi, TJ and Jafar, JD and Youhana, HM and Demidov, AM and Govor, LI},
  journal={Journal of Physics G: Nuclear Physics},
  volume={9},
  number={7},
  pages={779--795},
  year={1983}
}

@article{snelling1983gamma,
  title={Gamma-gamma directional correlation measurements in 146Nd following thermal-neutron capture},
  author={Snelling, DM and Hamilton, WD},
  journal={Journal of Physics G: Nuclear Physics},
  volume={9},
  number={1},
  pages={111--130},
  year={1983}
}

@article{ahmad1988coulomb,
  title={Coulomb excitation of 144, 146, 148, 150Nd},
  author={Ahmad, A and Bomar, G and Crowell, H and Hamilton, JH and Kawakami, H and Maguire, CF and Nettles, WG and Piercey, RB and Ramayya, AV and Soundranayagam, R and others},
  journal={Physical Review C},
  volume={37},
  number={5},
  pages={1836},
  year={1988},
  publisher={APS}
}

@article{pitz1990low,
  title={Low-energy photon scattering off 142,146,148,150 Nd: An investigation in the mass region of a nuclear shape transition},
  author={Pitz, HH and Heil, RD and Kneissl, U and Lindenstruth, S and Seemann, U and Stock, R and Wesselborg, C and Zilges, A and Von Brentano, P and Hoblit, SD and others},
  journal={Nuclear Physics A},
  volume={509},
  number={3},
  pages={587--604},
  year={1990},
  publisher={Elsevier}
}

@inproceedings{casten2007quantum,
  title={Quantum Phase Transitions in Finite Nuclei: Theoretical Concepts and Experimental Evidence},
  author={Casten, RF},
  booktitle={AIP Conference Proceedings},
  volume={899},
  number={1},
  pages={11--14},
  year={2007},
  organization={American Institute of Physics}
}

@article{ouhachi2018nuclear,
  title={Nuclear structure and decay properties of Nd isotopes},
  author={Ouhachi, M and Oudih, MR and Fellah, M and Allal, NH},
  journal={International Journal of Modern Physics E},
  volume={27},
  number={07},
  pages={1850059},
  year={2018},
  publisher={World Scientific}
}

@article{swain2025radial,
  title={Radial Sensitivity of the Nuclear Shell Structure at N= 92},
  author={Swain, RR and Anupam, A and Mohanty, P and Jena, KK and Agarwalla, SK and Sahu, BB},
  journal={Atom Indonesia},
  volume={51},
  number={1},
  pages={35--41},
  year={2025}
}

@article{gambhir1990relativistic,
  title={Relativistic mean field theory for finite nuclei},
  author={Gambhir, YK and Ring, P and Thimet, Ann},
  journal={Annals of Physics},
  volume={198},
  number={1},
  pages={132--179},
  year={1990},
  publisher={Elsevier}
}

@article{lalazissis1997new,
  title={New parametrization for the Lagrangian density of relativistic mean field theory},
  author={Lalazissis, GA and K{\"o}nig, J and Ring, P},
  journal={Physical Review C},
  volume={55},
  number={1},
  pages={540},
  year={1997},
  publisher={APS}
}

@article{swain2018nuclear,
  title={Nuclear structure and decay modes of Ra isotopes within an axially deformed relativistic mean field model},
  author={Swain, Rashmirekha and Patra, SK and Sahu, BB},
  journal={Chinese Physics C},
  volume={42},
  number={8},
  pages={084102},
  year={2018},
  publisher={Chinese Physical Society and the Institute of High Energy Physics of the~…}
}

@inproceedings{dash2023study,
  title={Study of nuclear structure and decay modes of Hassium},
  author={Dash, C and Tripathy, G and Naik, I and Sahu, BB},
  booktitle={AIP Conference Proceedings},
  volume={2901},
  number={1},
  pages={040020},
  year={2023},
  organization={AIP Publishing LLC}
}

@inproceedings{dash2021structural,
  title={Structural study of Es isotopes from $\alpha$-decay modes},
  author={Dash, C and Naik, I and Sahu, BB},
  booktitle={Proceedings of the DAE Symp. on Nucl. Phys},
  volume={65},
  pages={94},
  year={2021}
}

@inproceedings{dash2021shell,
  title={Shell closure at N\~{} 154 of Es element},
  author={Dash, C and Naik, I and Sahu, BB},
  booktitle={Proceedings of the DAE Symp. on Nucl. Phys},
  volume={65},
  pages={176},
  year={2021}
}

@inproceedings{dash2022probing,
  title={Probing the Shell Structures of 250-339 Hs108 Isotopes},
  author={Dash, C and Tripathy, G and Naik, I and Sahu, BB},
  booktitle={Proceedings of the DAE Symp. on Nucl. Phys},
  volume={66},
  pages={210},
  year={2022}
}

@inproceedings{dash2024nuclear,
  title={Nuclear Structure of Even Even Nd Isotopes C. Dash1, G. Tripathy2, A. Anupam2, I. Naik1, and BB Sahu2},
  author={Dash, C and Tripathy, G and Anupam, A and Naik, I and Sahu, BB},
  booktitle={Proceedings of the DAE Symp. on Nucl. Phys},
  volume={68},
  pages={243},
  year={2024}
}

@inproceedings{dash2023nuclear,
  title={Nuclear Structure of Even-Even 128-208Sm62 Isotopes},
  author={Dash, C and Naik, I and Tripathy, G and Sahu, BB},
  booktitle={Proceedings of the DAE-BRNS symposium on nuclear physics. V. 67},
  pages={2--2},
  year={2023}
}

@article{dash2026ground,
  title={The Ground State Aspects and the Impact of Shell Structures on the Stability of Es-Isotopes},
  author={Dash, C and Anupam, A and Naik, I and Sharma, BK and Sahu, BB},
  journal={arXiv preprint arXiv:2604.04624},
  year={2026}
}

@article{long2004new,
  title={New effective interactions in relativistic mean field theory with nonlinear terms and density-dependent meson-nucleon coupling},
  author={Long, Wenhui and Meng, Jie and Giai, Nguyen Van and Zhou, Shan-Gui},
  journal={Physical Review C—Nuclear Physics},
  volume={69},
  number={3},
  pages={034319},
  year={2004},
  publisher={APS}
}

@article{sharma1993rho,
  title={Rho meson coupling in the relativistic mean field theory and description of exotic nuclei},
  author={Sharma, MM and Nagarajan, MA and Ring, P},
  journal={Physics Letters B},
  volume={312},
  number={4},
  pages={377--381},
  year={1993},
  publisher={Elsevier}
}

@article{walecka1974theory,
  title={A theory of highly condensed matter},
  author={Walecka, John Dirk},
  journal={Annals of Physics},
  volume={83},
  number={2},
  pages={491--529},
  year={1974},
  publisher={Elsevier}
}

@article{serot1986adv,
  title={Adv in Nucl. Phys. vol. 16},
  author={Serot, BD and Walecka, JD and Negele, JW and Vogt, E},
  year={1986},
  publisher={Plenum, New York}
}

@article{boguta1977relativistic,
  title={Relativistic calculation of nuclear matter and the nuclear surface},
  author={Boguta, J and Bodmer, AR},
  journal={Nuclear Physics A},
  volume={292},
  number={3},
  pages={413--428},
  year={1977},
  publisher={Elsevier}
}

@article{sugahara1994relativistic,
  title={Relativistic mean-field theory for unstable nuclei with non-linear $\sigma$ and $\omega$ terms},
  author={Sugahara, YC and Toki, H},
  journal={Nuclear Physics A},
  volume={579},
  number={3-4},
  pages={557--572},
  year={1994},
  publisher={Elsevier}
}

@article{brockmann1992relativistic,
  title={Relativistic density-dependent Hartree approach for finite nuclei},
  author={Brockmann, R and Toki, H},
  journal={Physical review letters},
  volume={68},
  number={23},
  pages={3408},
  year={1992},
  publisher={APS}
}

@misc{serot1992relativistic,
  title={Relativistic nuclear many-body theory Recent Progress in Many-Body Theories ed TL Ainsworth et al},
  author={Serot, BD and Walecka, JD},
  year={1992},
  publisher={Springer}
}

@article{ring1996relativistic,
  title={Relativistic mean field theory in finite nuclei},
  author={Ring, Peter},
  journal={Progress in Particle and Nuclear Physics},
  volume={37},
  pages={193--263},
  year={1996},
  publisher={Elsevier}
}

@article{patra1991relativistic,
  title={Relativistic mean field study of light medium nuclei away from beta stability},
  author={Patra, SK and Praharaj, CR},
  journal={Physical Review C},
  volume={44},
  number={6},
  pages={2552},
  year={1991},
  publisher={APS}
}

@article{del2001pairing,
  title={Pairing properties in relativistic mean field models obtained from effective field theory},
  author={Del Estal, M and Centelles, M and Vinas, X and Patra, SK},
  journal={Physical Review C},
  volume={63},
  number={4},
  pages={044321},
  year={2001},
  publisher={APS}
}

@article{ring1997computer,
  title={Computer program for the relativistic mean field description of the ground state properties of even-even axially deformed nuclei},
  author={Ring, P and Gambhir, YK and Lalazissis, GA},
  journal={Computer physics communications},
  volume={105},
  number={1},
  pages={77--97},
  year={1997},
  publisher={Elsevier}
}

@techreport{nndc,
  title={{NNDC Data Services}},
  author={Tuli JK and Sonzogni A},
  year={2010},
  institution={\url{http://www.nndc.bnl.gov}, Brookhaven National Lab.(BNL), Upton, NY (United States)}
}

@article{moller2016nuclear,
  title={Nuclear ground-state masses and deformations: FRDM (2012)},
  author={M{\"o}ller, P and Sierk, Arnold John and Ichikawa, Takatoshi and Sagawa, Hiroyuki},
  journal={Atomic Data and Nuclear Data Tables},
  volume={109},
  pages={1--204},
  year={2016},
  publisher={Elsevier}
}

@article{krucken2002b,
  title={B (E 2) Values in N 150 d and the Critical Point Symmetry X (5)},
  author={Kr{\"u}cken, R and Albanna, B and Bialik, C and Casten, RF and Cooper, JR and Dewald, A and Zamfir, NV and Barton, CJ and Beausang, CW and Caprio, MA and others},
  journal={Physical review letters},
  volume={88},
  number={23},
  pages={232501},
  year={2002},
  publisher={APS}
}

@article{fossion20065,
  title={E (5), X (5), and prolate to oblate shape phase transitions in relativistic Hartree-Bogoliubov theory},
  author={Fossion, R and Bonatsos, Dennis and Lalazissis, GA},
  journal={Physical Review C—Nuclear Physics},
  volume={73},
  number={4},
  pages={044310},
  year={2006},
  publisher={APS}
}

@article{nikvsic2007microscopic,
  title={Microscopic description of nuclear quantum phase transitions},
  author={Nik{\v{s}}i{\'c}, Tamara and Vretenar, Dario and Lalazissis, GA and Ring, Peter},
  journal={Physical review letters},
  volume={99},
  number={9},
  pages={092502},
  year={2007},
  publisher={APS}
}

@article{robledo2008evolution,
  title={Evolution of nuclear shapes in medium mass isotopes from a microscopic perspective},
  author={Robledo, Luis Miguel and Rodr{\'\i}guez-Guzm{\'a}n, RR and Sarriguren, Pedro},
  journal={Physical Review C—Nuclear Physics},
  volume={78},
  number={3},
  pages={034314},
  year={2008},
  publisher={APS}
}

@article{rodriguez2008beyond,
  title={A beyond mean field analysis of the shape transition in the Neodymium isotopes},
  author={Rodriguez, Tom{\'a}s R and Egido, J Luis},
  journal={Physics Letters B},
  volume={663},
  number={1-2},
  pages={49--54},
  year={2008},
  publisher={Elsevier}
}

@article{cocolios2011early,
  title={Early onset of ground state deformation in neutron deficient polonium isotopes},
  author={Cocolios, Thomas Elias and Dexters, Wim and Seliverstov, MD and Andreyev, AN and Antalic, S and Barzakh, AE and Bastin, Beyhan and B{\"u}scher, Jeroen and Darby, IG and Fedorov, DV and others},
  journal={Physical review letters},
  volume={106},
  number={5},
  pages={052503},
  year={2011},
  publisher={APS}
}

@inproceedings{Zhang2020TowardAN,
  title={Toward a nuclear mass table with the continuum and deformation effects: even-even nuclei in the nuclear chart},
  author={Kaiyuan Zhang and Myung-Ki Cheoun and Yong-Beom Choi and Pooi Seong Chong and Jianmin Dong and Lisheng Geng and Eunja Ha and Xiaotao He and Changehoon Heo and Meng Chit Ho and Eun Jin In and Seonghyun Kim and Youngman Kim and Chang-Hwan Lee and Jenny Lee and Zhipan Li and Tianpeng Luo and Jie Meng and Myeong-Hwan Mun and Zhongming Niu and C. Pan and Panagiota Papakonstantinou and Xinle Shang and Caiwan Shen and Guofang Shen and Wei Sun and Xiang-Xiang Sun and Chi Kin Tam and Thaivayongnou and Chen Wang and Sau Hei Wong and Xuewei Xia and Yijun Yan and Ryan Wai-Yen Yeung and To Chung Yiu and Shuangquan Zhang and Wei Zhang and Shan-Gui Zhou},
  year={2020},
  url={https://api.semanticscholar.org/CorpusID:210839008}
}

@article{meng2003first,
  title={First order shape transition and critical point nuclei in Sm isotopes from relativistic mean field approach},
  author={Meng, J and Zhang, W and Zhou, S-G and Toki, H and Geng, LS},
  journal={arXiv preprint nucl-th/0312055},
  year={2003}
}

@article{zhang2016global,
  title={Global $\alpha$-decay study based on the mass table of the relativistic continuum Hartree-Bogoliubov theory},
  author={Zhang, Lin-Feng and Xia, Xue-Wei},
  journal={Chinese Physics C},
  volume={40},
  number={5},
  pages={054102},
  year={2016},
  publisher={IOP Publishing}
}

@article{rodriguez2010charge,
  title={Charge radii and structural evolution in Sr, Zr, and Mo isotopes},
  author={Rodr{\'\i}guez-Guzm{\'a}n, R and Sarriguren, Pedro and Robledo, Luis Miguel and Perez-Martin, S},
  journal={Physics Letters B},
  volume={691},
  number={4},
  pages={202--207},
  year={2010},
  publisher={Elsevier}
}

@article{kreim2014nuclear,
  title={Nuclear charge radii of potassium isotopes beyond N= 28},
  author={Kreim, Kim and Bissell, Mark L and Papuga, Jasna and Blaum, Klaus and De Rydt, Marieke and Ruiz, RF Garcia and Goriely, Stephane and Heylen, Hanne and Kowalska, Magdalena and Neugart, Rainer and others},
  journal={Physics Letters B},
  volume={731},
  pages={97--102},
  year={2014},
  publisher={Elsevier}
}

@article{gorges2019laser,
  title={Laser spectroscopy of neutron-rich tin isotopes: a discontinuity in charge radii across the N= 82 shell closure},
  author={Gorges, C and Rodr{\'\i}guez, LV and Balabanski, DL and Bissell, ML and Blaum, Klaus and Cheal, B and Garcia Ruiz, RF and Georgiev, G and Gins, W and Heylen, Hanne and others},
  journal={Physical review letters},
  volume={122},
  number={19},
  pages={192502},
  year={2019},
  publisher={APS}
}

@article{reponen2021evidence,
  title={Evidence of a sudden increase in the nuclear size of proton-rich silver-96},
  author={Reponen, M and de Groote, RP and Al Ayoubi, L and Beliuskina, O and Bissell, ML and Campbell, P and Ca{\~n}ete, L and Cheal, B and Chrysalidis, K and Delafosse, C and others},
  journal={Nature Communications},
  volume={12},
  number={1},
  pages={4596},
  year={2021},
  publisher={Nature Publishing Group UK London}
}

@article{day2021laser,
  title={Laser spectroscopy of neutron-rich Hg 207, 208 isotopes: Illuminating the kink and odd-even staggering in charge radii across the N= 126 shell closure},
  author={Day Goodacre, T and Afanasjev, AV and Barzakh, AE and Marsh, BA and Sels, S and Ring, P and Nakada, H and Andreyev, AN and Van Duppen, P and Althubiti, NA and others},
  journal={Physical review letters},
  volume={126},
  number={3},
  pages={032502},
  year={2021},
  publisher={APS}
}

@article{warbinek2024smooth,
  title={Smooth trends in fermium charge radii and the impact of shell effects},
  author={Warbinek, Jessica and Rickert, Elisabeth and Raeder, Sebastian and Albrecht-Sch{\"o}nzart, Thomas and Andelic, Brankica and Auler, Julian and Bally, Benjamin and Bender, Michael and Berndt, Sebastian and Block, Michael and others},
  journal={Nature},
  volume={634},
  number={8036},
  pages={1075--1079},
  year={2024},
  publisher={Nature Publishing Group UK London}
}

@article{malbrunot2022nuclear,
  title={Nuclear Charge Radii of the Nickel Isotopes Ni 58-68, 70},
  author={Malbrunot-Ettenauer, S and Kaufmann, S and Bacca, S and Barbieri, C and Billowes, J and Bissell, ML and Blaum, K and Cheal, B and Duguet, T and Ruiz, RF Garcia and others},
  journal={Physical Review Letters},
  volume={128},
  number={2},
  pages={022502},
  year={2022},
  publisher={APS}
}

@article{yang2023laser,
  title={Laser spectroscopy for the study of exotic nuclei},
  author={Yang, Xiaofei F and Wang, SJ and Wilkins, SG and Ruiz, RF Garcia},
  journal={Progress in Particle and Nuclear Physics},
  volume={129},
  pages={104005},
  year={2023},
  publisher={Elsevier}
}

@article{koszorus2021charge,
  title={Charge radii of exotic potassium isotopes challenge nuclear theory and the magic character of N= 32},
  author={Koszor{\'u}s, {\'A} and Yang, XF and Jiang, WG and Novario, SJ and Bai, SW and Billowes, J and Binnersley, CL and Bissell, ML and Cocolios, Thomas Elias and Cooper, BS and others},
  journal={Nature Physics},
  volume={17},
  number={4},
  pages={439--443},
  year={2021},
  publisher={Nature Publishing Group UK London}
}

@article{angeli2013table,
  title={Table of experimental nuclear ground state charge radii: An update},
  author={Angeli, Istv{\'a}n and Marinova, Krassimira Petrova},
  journal={Atomic Data and Nuclear Data Tables},
  volume={99},
  number={1},
  pages={69--95},
  year={2013},
  publisher={Elsevier}
}

@article{reinhard2016nuclear,
  title={Nuclear charge and neutron radii and nuclear matter: Trend analysis in Skyrme density-functional-theory approach},
  author={Reinhard, P-G and Nazarewicz, W},
  journal={Physical Review C},
  volume={93},
  number={5},
  pages={051303},
  year={2016},
  publisher={APS}
}

@article{hagen2015charge,
  title={Charge, neutron, and weak size of the atomic nucleus},
  author={Hagen, G and Ekstr{\"o}m, A and Forss{\'e}n, C and Jansen, GR and Nazarewicz, W and Papenbrock, T and Wendt, KA and Bacca, S and Barnea, N and Carlsson, B and others},
  journal={arXiv preprint arXiv:1509.07169},
  year={2015}
}

@techreport{gorgen2009study,
  title={Study of oblate nuclear shapes and shape coexistence in neutron-deficient rare earth isotopes},
  author={G{\"o}rgen, Andreas and Clement, Emmanuel and De France, G and Theisen, Ch and Wendt, A and Nyhus, HT and Sulignano, B and Tveten, GM and Larsen, AC and Girod, M and others},
  year={2009}
}

@article{rahaman2026shape,
  title={Shape transitions and ground-state properties of tungsten isotopes in covariant density functional theory: U Rahaman},
  author={Rahaman, Usuf},
  journal={Indian Journal of Physics},
  pages={1--14},
  year={2026},
  publisher={Springer}
}
%\end{multicols}

\end{document}